\documentclass[aps,pra,superscriptaddress,amsmath,amssymb,showpacs,twocolumn,notitlepage]{revtex4-1}
\usepackage{amssymb}
\usepackage{amsmath}
\usepackage{graphicx}
\usepackage{bm}
\usepackage{color}
\usepackage{subfigure}
\usepackage{multirow}
\usepackage{colortbl}
\usepackage{color}
\usepackage{soul}
\usepackage{footmisc}

\usepackage[bbgreekl]{mathbbol}
\usepackage{bbm}
\def\bbm[#1]{\mbox{\boldmath $#1$}}

\newcommand{\Q}{\dot{Q}}
\definecolor{ashgrey}{rgb}{0.7, 0.75, 0.71}

\graphicspath{{./Figures/}}

\makeatletter

    \def\CT@@do@color{%
      \global\let\CT@do@color\relax
            \@tempdima\wd\z@
            \advance\@tempdima\@tempdimb
            \advance\@tempdima\@tempdimc
    \advance\@tempdimb\tabcolsep
    \advance\@tempdimc\tabcolsep
    \advance\@tempdima2\tabcolsep
            \kern-\@tempdimb
            \leaders\vrule
            \@height\p@\@depth\p@
                    \hskip\@tempdima\@plus  1fill
            \kern-\@tempdimc
            \hskip-\wd\z@ \@plus -1fill }
    \makeatother

\newcolumntype{K}[1]{>{\centering\arraybackslash}p{#1}}

\begin{document}

\title{Excitation injector in an atomic chain: \\ long-ranged transport and efficiency amplification}

\author{Pierre Doyeux}
\affiliation{Laboratoire Charles Coulomb (L2C), UMR 5221 CNRS-Universit\'{e} de Montpellier, F- 34095 Montpellier, France}

\author{Riccardo Messina}
\affiliation{Laboratoire Charles Coulomb (L2C), UMR 5221 CNRS-Universit\'{e} de Montpellier, F- 34095 Montpellier, France}

\author{Bruno Leggio}
\affiliation{Laboratoire Charles Coulomb (L2C), UMR 5221 CNRS-Universit\'{e} de Montpellier, F- 34095 Montpellier, France}
\affiliation{Inria project Virtual Plants, CIRAD and INRA, F-34095 Montpellier, France}

\author{Mauro Antezza}
\affiliation{Laboratoire Charles Coulomb (L2C), UMR 5221 CNRS-Universit\'{e} de Montpellier, F- 34095 Montpellier, France}
\affiliation{Institut Universitaire de France, 1 rue Descartes, F-75231 Paris, France}

\begin{abstract}
We investigate the transport of energy in a linear chain of two-level quantum emitters (``atoms") weakly coupled to a blackbody radiation bath. We show that, simply by displacing one or more atoms from their regular-chain positions, the efficiency of the energy transport can be considerably amplified of at least one order of magnitude. Besides, in configurations providing an efficiency greater than $100\%$ , the distance between the two last atoms of the chain can be up to 20 times larger than the one in the regular chain, thus achieving a much longer-range energy transport. By performing both a stationary and time-dependent analysis, we ascribe this effect to an elementary block of three atoms, playing the role of excitation injector from the blackbody bath to the extraction site. By considering chains with up to 7 atoms, we also show that the amplification is robust and can be further enhanced up to $1400\%$.
\end{abstract}

\maketitle

\section{Introduction}

Energy transport in quantum systems is a field of research that has recently been drawing a lot of attention~\cite{dubi_colloquium_2011,ishizaki_quantum_2012,li_colloquium_2012}. The possibility of transporting energy efficiently in quantum systems is not only of fundamental interest, but also promises lots of potential technological applications, e.g. in quantum thermodynamics~\cite{leggio_quantum_2015,doyeux_quantum_2016}. As a matter of fact, it already has proven to be useful in numerous fields such as organic photovoltaic cells~\cite{menke_exciton_2014}, quantum information~\cite{wang_thermal_2007}, and more generally, nano-scale technologies~\cite{dubi_colloquium_2011}. In addition, the experimental observations of light-harvesting proteins, such as the FMO complex~\cite{brixner_two-dimensional_2005}, have also triggered a lot of interest~\cite{lloyd_symmetry-enhanced_2010,manzano_quantum_2013,celardo_cooperative_2014,feist_extraordinary_2015}. All these systems require energy to be transported within quantum systems over distances that might be relatively large, and with as little losses as possible. A deeper comprehension of the mechanism at the origin of energy transport in such systems is therefore essential. To this end, the investigation of systems composed of two-level quantum emitters provides an ideal framework~\cite{mohseni_environment-assisted_2008,chin_noise-assisted_2010,ishizaki_quantum_2012,dijkstra_role_2012,manzano_quantum_2013}.

One of the most challenging tasks remains to fully understand the role played by the environment in energy transport processes. This role has been investigated both in weak and strong coupling in many aspects~\cite{wang_nonequilibrium_2015}: quantum systems inside in a cavity~\cite{schachenmayer_cavity-enhanced_2015,feist_extraordinary_2015}, or placed close to metallic surfaces such as mirror~\cite{poddubny_collective_2015} or nano-spheres~\cite{gonzalez-ballestero_harvesting_2015}. The boundary conditions at the edges of linear systems have also been studied~\cite{bermudez_controlling_2013} as well as the activation of non-local effects stemming from environmental thermal bath~\cite{leggio_thermally_2015}.

In this latter work, the system under investigation is composed of a few two-level emitters embedded in blackblody radiation. The efficiency of the transport of incoherently-pumped excitations is studied with respect to the spatial distribution of the emitters. In particular, 2D and 3D configurations have been explored with the intention of mimicking the geometry of an FMO complex. Remarkably, in some cases, this efficiency can surpass the value of $100\%$ and even reach $300\%$, allowing one to extract three times more energy than the one pumped in. This is due to absorption of excitations during the transport process, which is triggered by non-local effects stemming from the presence of the thermal bath.

In this paper, we focus instead on the energy transport within a 1D system, that is a linear chain of a few two-level quantum emitters (from now on referred to as atoms) weakly interacting with blackbody radiation at an arbitrary temperature. Similarly to~\cite{leggio_thermally_2015}, the bath correlations provide the atoms with the possibility to (collectively) absorption excitations from the reservoir and to store them within the atomic system. We show that for specific geometrical configurations of the chain, atomic triplets can form an elementary block acting as an excitation injector, resulting in a remarkable enhancement of one order of magnitude of the efficiency. In other words, this latter can reach values greater that $1000\%$. In addition, these configurations realize energy transport over distances that are way larger than the one of the chain with regularly distributed atoms, which makes them particularly interesting for potential technological applications.

The paper is structured as follow: In Sec.~\ref{sec:PhysicalSystem} we introduce the model and the quantities used to perform the study of energy transport in our system. We present the main results of the paper in Sec.~\ref{sec:Long_range_transport_efficiency_amplification}, i.e. the occurrence of long-range efficiency amplification. Section \ref{sec:Discussion} is dedicated to the investigation, both at stationarity and during the dynamics, of a specific configuration of a 4-atom chain producing an amplified efficiency. Finally, in Sec.~\ref{sec:MoreAtoms}, we briefly investigate this amplification in linear chains composed of $N=5,6,7$ atoms.

\section{Physical system}
\label{sec:PhysicalSystem}

\begin{figure}[h!]
  \centering
  \includegraphics[width=.48\textwidth]{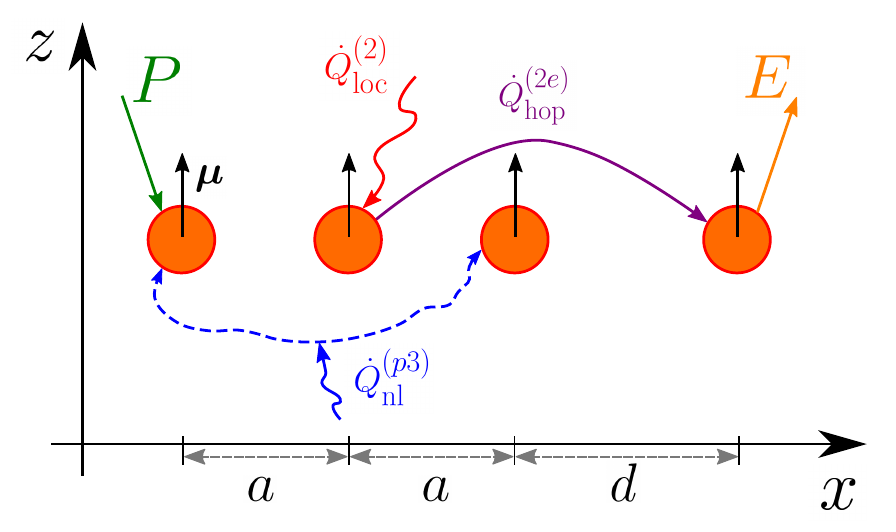}
  \caption{Linear chain of 4 atoms with dielectric dipole $\bbm[\mu]$ embedded in a thermal bath at $T$. Neighboring atoms are separated by a distance $a$, except the two last ones whose distance is $d$. Energy is pumped (extracted) within (from) the system with flux $P$ ($E$) applied on the first (last) atom of the chain. An example of each type of heat flux (local, non-local and hopping) has been represented on the figure.}
  \label{fig:Fig1}
\end{figure}

The open system under investigation, depicted in Fig.~\ref{fig:Fig1}, is a linear chain of $N$ identical two-level atoms of frequency $\omega$ which is embedded in an electromagnetic (EM) blackbody bath at temperature $T$. In order to investigate energy-transport efficiency along the chain, excitations are incoherently pumped into the system on one edge (the pumping site $p$), while extraction is performed on the other one (the extraction site $e$). The atoms composing the chain are thus labeled as $\{p, 2, ..., N-1, e\}$, and we introduce the Cartesian coordinates $\{x,y,z\}$ where the $x$-axis is defined as the one on which the atoms are aligned.

Following~\cite{leggio_thermally_2015}, the total Hamiltonian describing this system is
\begin{equation}
H_\mathrm{tot}=H_\mathrm{a}+H_\mathrm{B}+H_\mathrm{int},
\end{equation}
where $H_\mathrm{a}$ and $H_B$ are respectively the atomic and bath free Hamiltonians, and $H_\mathrm{int}$ is the interaction Hamiltonian. More precisely, being $|g_j\rangle$ ($|e_j\rangle$) the ground (excited) state of the $j$-th atom, and $\sigma_j^-=|g_j\rangle\langle e_j|$ ($\sigma_j^+=|e_j\rangle\langle g_j|$) the corresponding lowering (raising) operator, we have $H_\mathrm{a}~=~\hbar\omega\sum_{j=1}^{N} \sigma^+_j \sigma_j^-$. Besides, assuming that each atom has the same dielectric dipole moment operator $\bbm[\mu]$ and within the dipole approximation, the atom-bath interaction is given by $H_\mathrm{int}=-\sum_{j=1}^N \bbm[\mu] \cdot \bold{E}(\bold{r}_j)$ where $\bold{E}(\bold{r}_j)$ is the electric field at the position $\bold{r}_j$ of the $j$-th atom.

\subsection{Master equation}
Within the weak-coupling limit, the Born-Markov approximation~\cite{breuer_theory_2002} allows us to describe the time evolution of the reduced atomic density matrix $\rho(t)$ with the following quantum Markovian master equation~\cite{bellomo_creation_2013,leggio_thermally_2015}
\begin{equation}\label{ME}
\dot{\rho}=-\frac{i}{\hbar}\big[H_{\mathrm{sys}},\rho\big]+D_{\mathrm{loc}}[\rho]+ D_{\mathrm{nl}}[\rho]+D_{\mathrm{in}}[\rho]+D_{\mathrm{out}}[\rho],
\end{equation}
where here and in the following the explicit time dependence of $\rho$ and $\rho$-dependent quantities will be dropped for simplicity. The effective Hamiltonian $H_\mathrm{sys} = H_\mathrm{a} + H_\Lambda$, where
\begin{equation}
H_\Lambda =  \sum_{j<k}^N H_\Lambda^{(jk)}=\hbar\sum_{j<k}^N\Lambda_{jk}\big(\sigma^+_j\sigma_k^-+\sigma^+_k\sigma_j^-),
\end{equation}
characterizes the unitary evolution of the atomic system. The coefficient $\Lambda_{jk}$ is the interaction strength between induced dipoles of the transitions of atoms $j$ and $k$, which is mediated by the EM field, and is given by
\begin{equation}\begin{split}
\Lambda_{jk}&=-\frac{3}{4}\gamma_0\Big[2(\hat{\bbm[\mu]}\cdot\hat{\mathbf{r}}_{jk})^2f(\tilde{r}_{jk})+\Big(1-(\hat{\bbm[\mu]}\cdot\hat{\mathbf{r}}_{jk})^2\Big)g(\tilde{r}_{jk})\Big],\label{lambda}\\
f(x)&=\frac{\cos{x}+x\sin{x}}{x^3},\quad g(x)=\frac{(x^2-1)\cos{x-x\sin{x}}}{x^3},
\end{split}\end{equation}
where the coefficient $\gamma_0= |\bbm[\mu]|^2\omega^3/3\pi\epsilon_0\hbar c^3$ is the rate of spontaneous emission in vacuum. We have also introduced $\hat{\bbm[\mu]}=\frac{\bbm[\mu]}{|\bbm[\mu]|}$, $\bold{\hat{r}}_{jk}= \frac{\bold{r}_{jk}}{|\bold{r}_{jk}|}$ where $\bold{r}_{jk}=\bold{r}_j-\bold{r}_{k}$ and $\tilde{r}_{jk} = \frac{\omega}{c}|\bold{r}_{jk}|$. It is worth stressing that $\Lambda_{jk}$ dramatically increases when the distance between the two atoms is smaller than $c/\omega$.

The non-unitary interaction between the atomic system and the bath is described by the so-called dissipators. Firstly, each atomic transition exchanges excitations with the thermal bath by emitting/absorbing a photon of frequency $\omega$. The local dissipators capture these interactions
\begin{equation}\label{local}
\begin{split}
D_{\mathrm{loc}}[\rho]&=\sum_j D_\mathrm{loc}^{(j)}[\rho] \\
&=\sum_j\bigg\{n\gamma_0\Big(\sigma^+_j\rho\sigma^-_j-\frac{1}{2}\{\sigma^-_j\sigma^+_j, \rho\}\Big)\\
&+(1+n)\gamma_0\Big(\sigma^-_j\rho\sigma^+_j-\frac{1}{2}\{\sigma^+_j\sigma^-_j, \rho\}\Big)\bigg\},
\end{split}
\end{equation}
where $n=\big[\text{exp}(\hbar\omega/k_BT)-1\big]^{-1}$ is the average number of photons with energy $\hbar \omega$ when the bath is at temperature $T$ and with $\{\cdot,\cdot\}$ denoting an anticommutator.

In addition to a local dissipator, involving atoms separately, we should also take into account that each couple of atoms may interact with the EM field with a collective coherent behavior: the two atoms behave as a single entity, and they trade the same amount of energy with the environmental bath, both of them absorbing or emitting energy in a unique process. This non-local dissipation phenomenon is triggered by the auto-correlation functions of the EM field which allow resonant atomic transitions to cooperate. The corresponding non-local dissipator reads
\begin{equation}\label{nonlocal}
\begin{split}
D_{\mathrm{nl}}[\rho]&=\sum_{j< k}D_{\mathrm{nl}}^{(jk)}[\rho] \\
&=\sum_{j< k}\bigg\{n\gamma_{jk}\Big(\sigma^+_j\rho\sigma^-_k-\frac{1}{2}\{\sigma^-_k\sigma^+_j, \rho\}\Big)\\
&+(1+n)\gamma_{jk}\Big(\sigma^-_j\rho\sigma^+_k-\frac{1}{2}\{\sigma^+_k\sigma^-_j, \rho\}\Big)+\text{h.c.}\bigg\}.
\end{split}
\end{equation}
where $\gamma_{jk}$ is the rate of spontaneous emission of the atomic pair $(j,k)$ which is obtained as
\begin{equation}
\gamma_{jk}=\gamma_0\sum_{l=1,2,3}\Big([\hat{\bbm[\mu]}]_l\Big)^2\alpha_{jk}^{(l)},
\end{equation}
$l$ denoting the three Cartesian components, and where we used
\begin{eqnarray}
\alpha_{jk}^{(1)}&=&\frac{3}{\tilde{r}_{jk}^3}\Big(\sin{\tilde{r}_{jk}}-\tilde{r}_{jk}\cos{\tilde{r}_{jk}}\Big),\\
\alpha_{jk}^{(2)}&=&\alpha_{jk}^{(3)}=\frac{3}{2\tilde{r}_{jk}^3}\Big(\tilde{r}_{jk} \cos{\tilde{r}_{jk}}+(\tilde{r}_{jk}^2-1)\sin{\tilde{r}_{jk}}\Big).
\end{eqnarray}

Under the sole action of the local and non-local dissipators, the system would reach thermal equilibrium with the bath at temperature $T$. However, the presence of incoherent pumping and extraction perturbs the system, preventing it from reaching its Gibbs state. The pumping performed on $p$ is described by
\begin{equation}\label{Din}
D_{\mathrm{in}}[\rho]=\gamma_{\mathrm{in}}\left(\sigma^+_p\rho\sigma^-_p-\frac{1}{2}\{\sigma^-_p\sigma^+_p,\rho\}\right),
\end{equation}
where $\gamma_\mathrm{in}$ is the pumping rate. The dissipator corresponding to the extraction occurring on $e$ with rate $\gamma_\mathrm{out}$ is
\begin{equation}\label{Dout}
D_{\mathrm{out}}[\rho]=\gamma_{\mathrm{out}}\left(\sigma^-_e\rho\sigma^+_e-\frac{1}{2}\{\sigma^+_e\sigma^-_e,\rho\}\right).
\end{equation}

\subsection{Heat fluxes and efficiency}

In order to understand the energy-transport properties of our system, it is natural to identify the different channels through which the atomic system can either absorb, emit or transmit energy. Before entering into details, let us precise that the energy exchanges occurring in our system are heat fluxes as defined in the framework of quantum thermodynamics~\cite{balian_microphysics_1991,gemmer_quantum_2004}. No work is involved since $\partial H_\mathrm{sys}/\partial t=0$.

Let us start with the so-called hopping heat fluxes. These occur between two atomic resonant transitions and represent energy exchanged between two atoms without modifying the energy content of the environment. Therefore, these fluxes do not change the energy of the global atomic system but rather describe trades of excitations between its different subparts. More specifically, they stem from the field-induced dipole--dipole interactions and, for the atom $j$, are defined as
\begin{equation}
\Q_\mathrm{hop}^{(jk)} = -\frac{i}{\hbar}\,\text{Tr} \big(H_\mathrm{a}^{(j)}\big[H_\Lambda^{(jk)},\rho\big]\big),
\label{eq:HoppingFluxes1}
\end{equation}
where $j$ and $k$ ($j\mathop{\neq}k$) index two atoms and with $H_\mathrm{a}^{(j)}~=~\hbar\omega\sigma^+_j\sigma_j$. With this definition, having $\dot{Q}_\mathrm{hop}^{(jk)}>0$ means that atom $j$ is absorbing energy at the expense of $k$. Moreover, note that $\Q_\mathrm{hop}^{(kj)}=-\Q_\mathrm{hop}^{(jk)}$. It is worth stressing that Eq.~\eqref{eq:HoppingFluxes1} can be rewritten, after straightforward calculations, as
\begin{equation}
\dot{Q}_\mathrm{hop}^{(jk)}=-2\,\hbar\omega\Lambda_{jk}\text{Im}(c^{(jk)}),
\label{eq:HoppingFluxes2}
\end{equation}
where $c^{(jk)}=\langle\sigma_j^+\sigma_k^-\rangle$ is the coherence between atoms $j$ and $k$. Equation \eqref{eq:HoppingFluxes2} encapsulates two essential features of the hopping fluxes. Firstly, they depend on the amplitude of the dipole--dipole interaction $\Lambda_{jk}$, notably inheriting its spatial dependence. Secondly, the existence and the strength of the hopping fluxes directly depend on the presence of non-real coherences between the two atoms. Note that this term of energy hopping between sites is often the only one considered in standard transport models allowing energy to be transported along atomic systems~\cite{feist_extraordinary_2015,schachenmayer_cavity-enhanced_2015}.

We now turn to the fluxes stemming from the dissipators, starting with the local ones. They describe the exchanges of excitations between an atomic transition and the modes of the EM bath at the same frequency:
\begin{equation}
\Q_\mathrm{loc}^{(j)}=\text{Tr}\big(H_\mathrm{a}D^{(j)}_\mathrm{loc}[\rho]\big),
\label{eq:LocalFlux}
\end{equation}
where $\Q_\mathrm{loc}^{(j)} > 0$ ($<0$) means that the transition is absorbing (emitting) energy from (to) the environment.

Similarly, the non-local dissipators produce heat fluxes between an atomic pair and the modes of the environmental field. Focusing for example on the couple $(j,k)$, the energy absorbed (emitted) by atom $j$ reads
\begin{equation}
\Q_\mathrm{nl}^{(jk)}=\text{Tr}\big(H^{(j)}_a D^{(jk)}_\mathrm{nl} [\rho]\big).
\label{eq:NonLocFlux}
\end{equation}
In the case $\Q_\mathrm{nl}^{(jk)}>0$ ($<0$), the atoms $j$ and $k$ both absorb (emit) an equal amount of energy $\Q_\mathrm{nl}^{(jk)}$ from (to) the EM field. Straightforward calculations lead to the following expression of the non-local heat fluxes:
\begin{equation}
\Q^{(jk)}_\mathrm{nl}=-\hbar\omega\gamma_{jk}\text{Re}(c^{(jk)}).
\label{eq:NonLocFlux2}
\end{equation}
Thus, similarly to the hopping fluxes, the non-local ones can only exist in the presence of coherences. It is precisely the presence of these terms, usually neglected in models of quantum energy transport, that triggers remarkable effects on the transport efficiency, as will be clarified in the next sections.

Finally, the heat flux of excitations injected into $p$ is
\begin{equation}
P=\mathrm{Tr}\big(H_\mathrm{a} D_{\mathrm{in}}[\rho]\big).
\label{P}
\end{equation}
Note that $P\geq0$ since the energy of the atomic system is always increased by this flux.

On the other hand, the extraction lowers the energy of the atomic system. Thus, in order to have a positive quantity, we define
\begin{equation}
E=-\mathrm{Tr}\big(H_\mathrm{a} D_{\mathrm{out}}[\rho]\big),
\label{E}
\end{equation}
which corresponds to the energy extracted from $e$, so that $E\geq0$.

Following [21], we use the quantities introduced in Eqs.~\eqref{P} and \eqref{E} to investigate the energy-transport efficiency along the chain, which we define as
\begin{equation}
\chi=\frac{E-E_0}{P},
\label{chi}
\end{equation}
where $E_0$ is the field thermal energy extracted from $e$ in the absence of pumping, i.e., in the limit $\gamma_\mathrm{int}\rightarrow0$. Note that, although $\gamma_\mathrm{int}$ does not \textit{explicitly} appear in Eq.~(18), $E$ is a function of $\gamma_\mathrm{int}$ due to the dependence of the atomic state $\rho$ on the pumping rate.

In particular, having $\chi=1$ means that the amount of energy pumped into $p$ equals the extra energy extracted in the presence of pumping, suggesting that the excitations have been transported throughout the chain without any loss or gain. In general $\chi$ is a function of time because of the $t$-dependence of $\rho$. In what follows we will study both the stationary efficiency $\chi_{\text{stat}}$, obtained in the limit $t\rightarrow \infty$, and the dynamic behavior of the efficiency.

\section{Long-range transport and efficiency amplification}
\label{sec:Long_range_transport_efficiency_amplification}

The issue we address here is the robustness of energy-transport efficiency with respect to the displacement of the extraction site, the last atom of the chain in our case. To this aim we start with the 4-atom configuration ($N=4$) represented in Fig.~\ref{fig:Fig1}. In this simulation,  the atomic frequency is $\omega=1\times10^{14}\,\text{rad}\cdot\text{s}^{-1}$ and the dipoles are oriented orthogonally to the chain with magnitude $|\bbm[\mu]|=10^{-30}\,\text{C}\cdot \text{m}$. The rates of pumping and extraction are $\gamma_\mathrm{in}=10^{-3}\gamma_0$ and $\gamma_\mathrm{out}=10^{2}\gamma_0$. These values have been used for instance in [21] to simulate models analogous to FMO-complexes. All the numerical results have been obtained using the open-source package QuTiP~\cite{johansson_qutip:_2012,johansson_qutip_2013}.

\begin{figure}[h!]
  \centering
  \includegraphics[width=.48\textwidth]{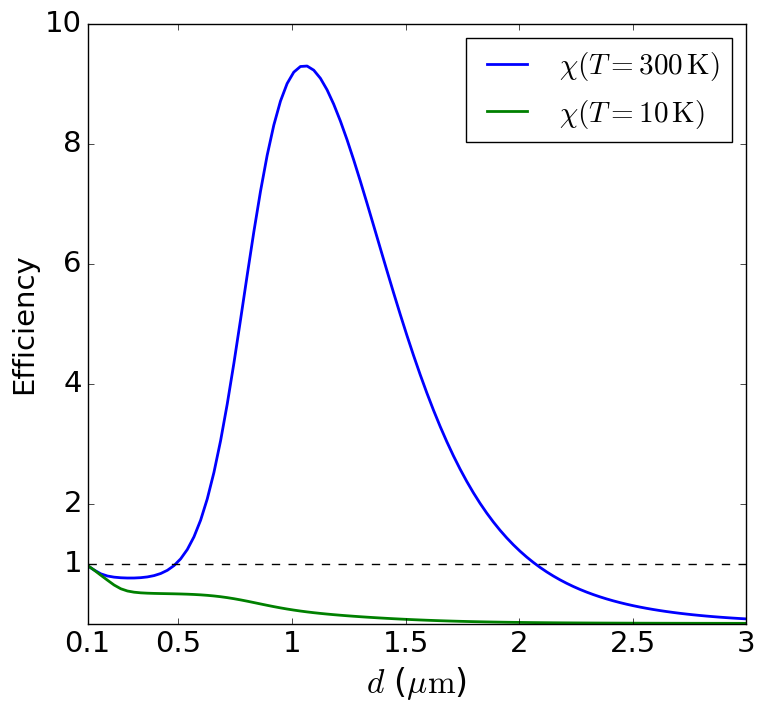}
  \caption{Efficiency as a function of the distance between atom $e$ and the rest of the chain at $T=10\,$K (green line) and $T=300\,$K (blue line).}
  \label{fig:Fig2}
\end{figure}

We start our analysis with Fig.~\ref{fig:Fig2}, which shows the stationary efficiency as atom $e$ is moved away from the rest of the chain for two different temperatures. The regular configuration we take as a reference has $d=a=0.1\,\mu$m.

The green curve corresponds to $T=10\,$K at which the hopping fluxes dominate over the local and non-local ones. Indeed, the latter explicitly depend on the temperature because of the presence of the thermal factor $n$ in the expression of the local and non-local dissipators. As a consequence, since $n\to0$ as $T\to0\,$K, lowering the environmental temperature tends to suppress the interactions between the atomic system and the EM thermal bath. On the other hand, the interaction strength $\Lambda_{jk}$ characterizing the dipole--dipole interaction between the atomic couple $(j,k)$ is temperature-independent. Consequently, the hopping fluxes depend on $T$ only indirectly through  the state $\rho$, and are much less affected by a change of the environmental temperature than the local and non-local ones. In particular, unlike the dissipative fluxes, the hopping ones do not vanish in the limit $T\to0\,$K.

This explains why, in the case $T=10\,$K, the interactions between the atomic system and the EM bath are limited. Thus, the excitations pumped into $p$ are transmitted to atom $2$ via hopping without almost any loss, and so forth until they reach the extraction site $e$. As a consequence, the regular chain leads to an efficiency close to $\chi (T\mathop{=}10\,\text{K})\simeq100\%$. As $d$ increases, the interaction strength $\Lambda_{3e}$ between atoms 3 and $e$ decreases, inducing a diminution of the hopping flux between them: the hopping-transmission chain weakens, resulting in an efficiency collapsing to $0$.

The initial scenario ($d\simeq0.1\,\mu$m) for the high-$T$ case is quite similar, with $\chi\sim 1$ for a regular chain. However, contrarily to the low-temperature scenario, pulling away the extraction atom at $T=300\,$K results in a remarkable efficiency amplification. In particular, efficiency reaches a maximum value greater than $900\%$. This means that the increase of extracted energy from $e$ due to pumping is much larger than the energy injected into $p$. This suggests that the atomic system draws the additional excitations from the environmental thermal bath. A detailed investigation of this phenomenon will be the object of Sec.~\ref{sec:Discussion}.

\begin{figure}[h!]
  \centering
  \includegraphics[width=.48\textwidth]{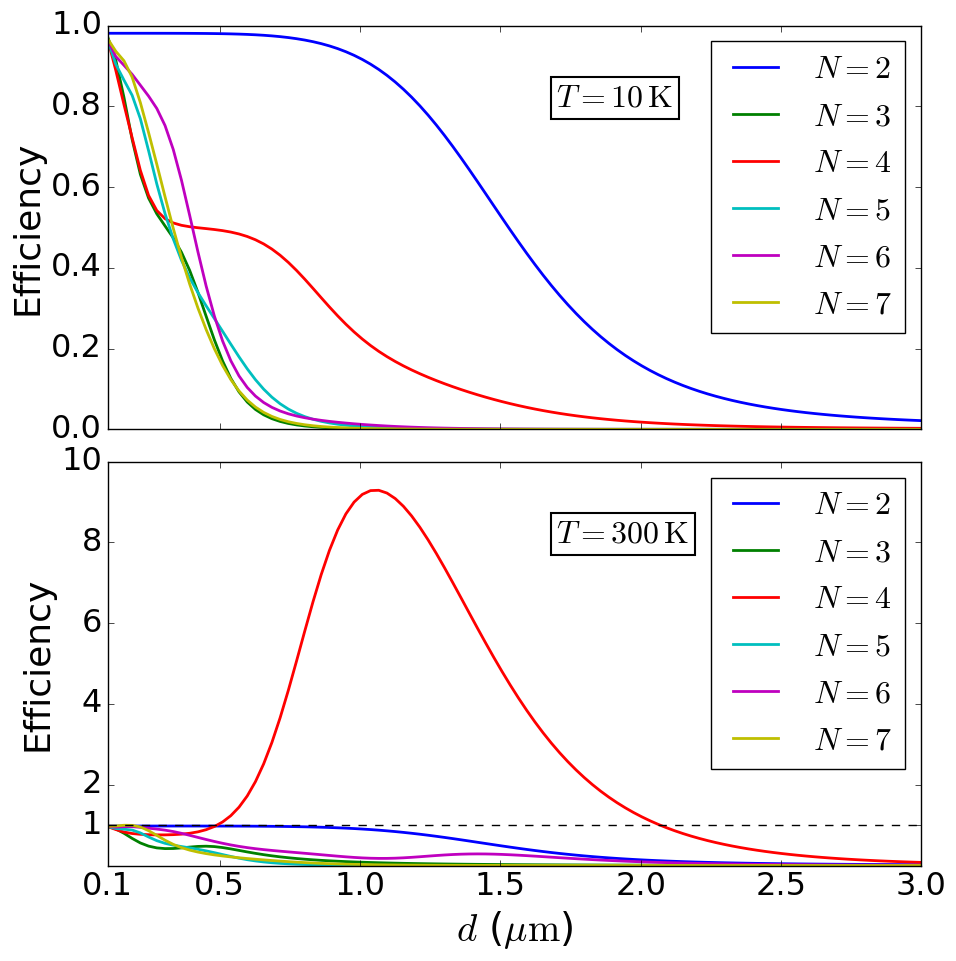}
  \caption{Efficiency as a function of the distance between atom $e$ and atom $N-1$ for different values of $N$ (total number of atoms composing the chain). The upper plot is realized at $T=10\,$K and the lower one at $T=300\,$K.}
  \label{fig:Fig3}
\end{figure}

This dramatic efficiency amplification, of almost one order of magnitude, is the main result of this paper. We have shown that despite the simple 1D geometry of the system, a clever manipulation of thermal and geometrical parameters can be exploited to actively tune the amount of energy transported within the chain. Compared to previous works concerning 2D and 3D geometries~\cite{leggio_thermally_2015}, our 1D chain is able to produce an effiency at least three times greater. Moreover, the effect we highlight is achievable in a wide range of geometrical configurations. More specifically, while for $T=10\,$K the only chain realizing $\chi\simeq1$ is the regular one, for $T=300\,$K an efficiency larger than 1 is produced within the entire interval $d\in\left[0.5, 2\right]\,\mu$m, including distances between the two last atoms up to 20 times greater than the regular-chain spacing. This proves that this long-range amplified energy transport is robust with respect to small displacements of atom $e$, and results indeed from a combination of thermal and geometrical features.

In order to understand the behavior of the long-range efficiency amplification with respect to the number of atoms, we plot in Fig.~\ref{fig:Fig3} the efficiency against $d$ for $N=2,\dots,7$. For $T=10\,$K the behavior we observe for any $N$ is qualitatively the same as the one we already had for $N=4$: an efficiency starting from 1 and collapsing to 0 for large distances. For $T=300\,$K, on the contrary, we clearly see that the case N = 4, which is the same as Fig. 2, is the only one showing a strong amplification of the efficiency. For this reason, in the following, we focus our study on this specific configuration, leaving for Sec.~\ref{sec:MoreAtoms} a discussion of the case $N>4$.

\begin{figure*}[t!]
  \begin{center}
    \includegraphics[width=.96\textwidth]{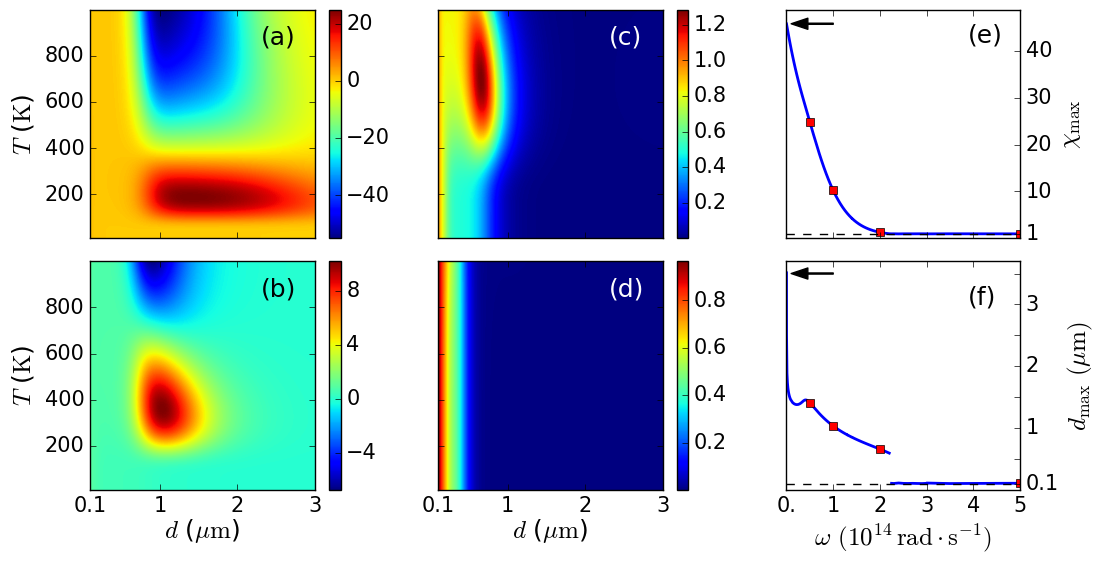}
    \caption{Panels (a)-(d): Efficiency as a function of the distance between atom $e$ and the rest of the chain and of the bath temperature for different frequencies: (a): $\omega=0.5\times 10^{14}\,\text{rad}\cdot\text{s}^{-1}$, (b): $\omega=1\times 10^{14}\,\text{rad}\cdot\text{s}^{-1}$, (c): $\omega=2\times 10^{14}\,\text{rad}\cdot\text{s}^{-1}$, (d): $\omega=5\times 10^{14}\,\text{rad}\cdot\text{s}^{-1}$. Maximum efficiency (panel (e)) and distance at which this maximum is realized (panel (f)) as a function of the frequency. The maximization has been performed with respect to the parameters $d\in[0.1,3]\,\mu$m and $T\in[10,1000]\,$K. The black arrows in panels (e) and (f) indicate the asymptotic limit of respectively $\chi_\text{max}$ and $d_\text{max}$ at low frequencies.}
    \label{fig:Fig4}
  \end{center}
\end{figure*}

We start our investigation with Fig.~\ref{fig:Fig4}, where we analyze the dependence of the efficiency amplification on different parameters. More specifically, panels (a)-(d) are density plots of $\chi$ as a function of $d$ and of the temperature $T$ for different transition frequencies.

The first feature to point out is that panels (a)-(c) show efficiency-amplified regions ($\chi > 1$), witnessing a non-monotonic behavior of $\chi$ with respect to both parameters $d$ and $T$, absent in panel (d). Note that in some regions, the efficiency takes negative values, meaning that pumping excitations into the system reduces the energy extracted from the chain ($E<E_0$). Moreover, we must stress that the values taken by the efficiency are very different for each panel, showing a strong dependence of this effect on the atomic transition frequency. In order to have a deeper insight on this dependence, panel (e) shows the maximum efficiency with respect to both variables $d \in [0.1,3]\,\mu\text{m}$ and $T\in[10,1000]\,$K as a function of $\omega$. Remarkably, at frequencies around $\omega\simeq 0.1\times 10^{14}\,\text{rad}\cdot\text{s}^{-1}$, the efficiency reaches $\chi_\text{max}\simeq 40$. Then it decreases when increasing $\omega$, until reaching values $\leq 1$, meaning that the efficiency amplification is no longer realized at frequencies $\geq 2\times 10^{14}\,\mathrm{rad}\cdot\mathrm{s}^{-1}$.

In the asymptotic limit of low frequencies, indicated by black arrows in panels (e) and (f) of Fig.~\ref{fig:Fig4}, the efficiency reaches a plateau at $\chi_\text{max}\simeq 45$ for $\omega\leq0.01\times10^{14}\,\text{rad}\cdot\text{s}^{-1}$ \cite{remark}.

\begin{figure}[t!]
  \begin{center}
    \includegraphics[width=.96\linewidth]{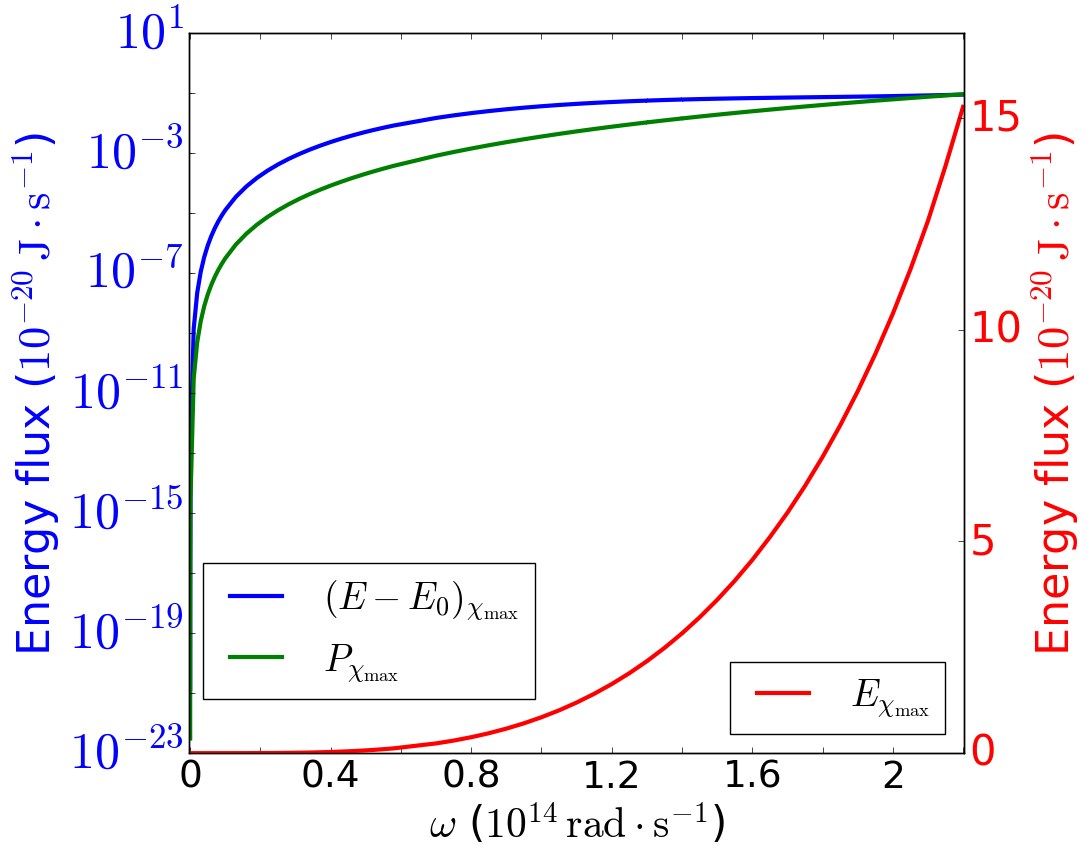}
    \caption{Left vertical scale: difference
$(E-E_0)_{\chi_\text{max}}$ of the energy
extracted from the chain between the scenarios with and without pumping (blue
solid line) and $P_{\chi_\text{max}}$ of the energy pumped
into the chain (solid green line). Right vertical scale: $E_{\chi_\text{max}}$
of the energy extracted from the chain in the pumping case. All the quantities
are plotted as a function of the frequency and correspond to the maximum of
efficiency for every frequency,
i.e. $(E-E_0)_{\chi_\text{max}}/P_{\chi_\text{max}}$ results
in $\chi_\text{max}$ plotted in panel (e) of Fig.~\ref{fig:Fig4}.}
    \label{fig:EE0P}
  \end{center}
\end{figure}

To have a deeper insight on the pumping and extraction fluxes entering into play, Fig.~\ref{fig:EE0P} pictures the quantities from which $\chi_\text{max}$ of Fig.~\ref{fig:Fig4} is obtained in the region where $\chi_\text{max}>1$. As one can see, we have $P_\text{max} \ll E_\text{max}$, and increasing the frequency results in increasing both $P_\text{max}$ and $E_\text{max}$. These two features stem respectively from the facts that $\gamma_\text{in} \ll \gamma_\text{out}$ and that both these coefficients depend on $\gamma_0$.

Concerning the position of the maximum of efficiency in panels (a)-(d) of Fig.~\ref{fig:Fig4}, one can notice two important features when increasing the frequency: the temperature at which $\chi_\mathrm{max}$ is reached increases as well, and the distance of the last atom providing this maximum approaches the one of the regular-chain configuration. To study more in detail this phenomenon, we plot in panel (f) the displacement of atom $e$ at which $\chi_\text{max}$ is reached as a function of the frequency. Clearly, the distance $d_\text{max}$ decreases when increasing $\omega$. The discontinuity at $\omega\simeq2.3\times 10^{14}\,\mathrm{rad}\cdot\mathrm{s}^{-1}$ witnesses the transition from the long-range energy-transport configuration, in which the efficiency is amplified, to the regular-chain one, where $\chi_\mathrm{max}\leq 1$. In other words, while for $\omega\gtrsim2.3\times 10^{14}\,\mathrm{rad}\cdot\mathrm{s}^{-1}$ the best efficiency ($\leq$ 1) is always realized by the regular configuration ($d=a$), for any lower frequencies we observe an optimized distance $d_\text{max}$ producing a long-range efficiency amplification.

\section{Discussion}
\label{sec:Discussion}
\subsection{Steady-state analysis}
\label{SteadystateAnalysis}

According to Eq.~\eqref{chi}, the efficiency characterizes the difference in the energy extracted from the chain between the scenarios with and without pumping. Therefore, in order to understand the origin of the efficiency amplification observed in the previous section, it is natural to investigate the effect of the pumping on the energy exchanges occurring both inside the atomic system and between the atoms and the bath. We start this subsection by introducing the quantities we use to study these effects. These are valid at any time $t$, but we will start our investigation by focusing on the steady state of the atomic system, i.e. the state reached for $t\to+\infty$. Moreover, for the simulations performed in both subsections \ref{SteadystateAnalysis} and \ref{Dynamics}, the atomic frequency is $\omega=1\,\times10^{14}\text{rad}\cdot\text{s}^{-1}$ and the temperature is fixed at $T=361\,$K, the one corresponding to the maximum observed in Fig.~\ref{fig:Fig4}(b).

In order to characterize the effect of the pumping on the heat fluxes, we define the differences
\begin{equation}\begin{split}
  \Delta \Q_\text{loc}^{(j)}&=\Q_\text{loc}^{(j)}-\Q_{\text{loc},0}^{(j)},\\
  \Delta \Q_\varphi^{(jk)}&=\Q_\varphi^{(jk)}-\Q_{\varphi,0}^{(jk)},
  \end{split}
  \label{eq:DeltaFlux}
\end{equation}
where the index $\varphi\in\{\text{nl}, \text{hop}\}$ according to the
type of flux under investigation. The label $0$ corresponds to the scenario in
which no pumping is performed. We also define, for a given atom $j$,
\begin{equation}
  \Delta \Q^{(j)}_\varphi=\sum_{k\neq j}\Delta \Q_\varphi^{(jk)},
  \label{eq:DeltaFluxContribution}
\end{equation}
which describes the total contribution of the type of flux under consideration involving atom $j$ (once again $\varphi\in\{\text{nl}, \text{hop}\}$).

\begin{figure}[b!]
  \centering
  \includegraphics[width=.48\textwidth]{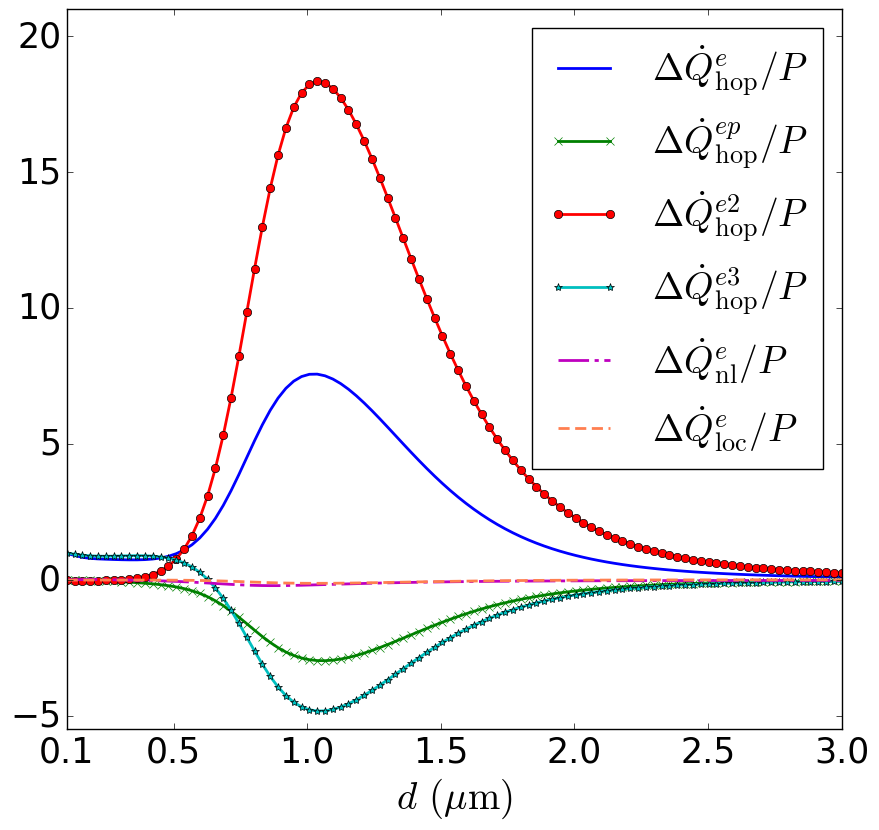}
  \caption{Difference of the heat fluxes involving atom $e$ between the scenarios with and without pumping at steady state divided by $P$. Each type of heat flux is represented: $\Delta Q^{(e)}_\text{hop}$ (solid blue line), $\Delta Q^{(e)}_\text{nl}$ (long-dashed purple line), $\Delta Q^{(e)}_\text{loc}$ (dashed orange line). The contribution of each atom to $\Delta Q^{(e)}_\text{hop}$ is also plotted: $\Delta Q^{(ep)}_\text{hop}$ (solid green line with crosses), $\Delta Q^{(e2)}_\text{hop}$ (solid red line with circles), $\Delta Q^{(e3)}_\text{hop}$ (solid cyan line with stars).}
\label{fig:Fig5}
\end{figure}

To track the root of the efficiency-amplification mechanism, we start our analysis by investigating the effect of pumping on each kind of flux involving atom $e$. Figure~\ref{fig:Fig5} shows $\Delta\Q^{(e)}_\text{loc}$, $\Delta\Q^{(e)}_\text{nl}$ and $\Delta\Q^{(e)}_\text{hop}$ as a function of $d$ at stationarity (the extraction flux is not shown). As one can see, the hopping contribution has a behavior extremely similar to the one of the efficiency (see Fig.~\ref{fig:Fig2}), while the local and non-local contributions are small with respect to the hopping one. In particular, having $\Delta \Q^{(e)}_\mathrm{hop}>0$ hints that the efficiency amplification arises from the enhancement (due to pumping) of the energy received by $e$ through hopping fluxes from the rest of the chain.

More specifically, unveiling the contribution of each atom to $\Delta \Q^{(e)}_\mathrm{hop}$ provides further insight. Indeed, the hopping flux between atoms $2$ and $e$ is clearly the most affected by the displacement of the latter. In particular, in the interval $d\in[0.5, 2]\,\mu$m, that is when efficiency amplification is realized, we have $\Delta \Q_\mathrm{hop}^{(e2)}>0$, meaning that the energy flowing from atom $2$ to $e$ is increased in the presence of pumping. The opposite behavior occurs for the couples $(e,p)$ and $(e,3)$, the variation being less pronounced than the one of $\Delta \Q_\mathrm{hop}^{(e2)}$, which explains why the resulting hopping contribution verifies $\Delta \Q_\mathrm{hop}^{(e)}>0$. Therefore, the efficiency amplification stems on the particularly enhanced hopping flux between atoms $2$ and $e$. Note that in the regular-chain configuration ($d=0.1\,\mu$m), atom $e$ receives the additional energy due to pumping only by hopping from its closest neighbor, that is atom $3$.

\begin{figure}[h!]
  \centering
  \includegraphics[width=.48\textwidth]{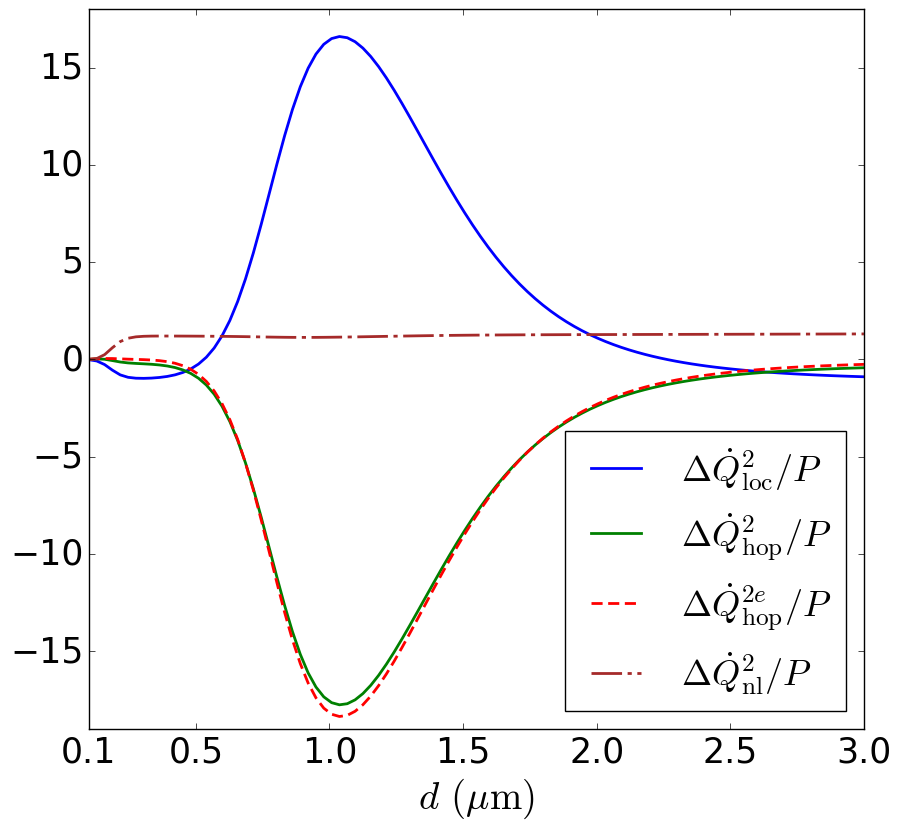}
  \caption{Difference of the heat fluxes involving atom $2$ between the pumping and no-pumping cases normalized by $P$ at steady state. Each type of flux is plotted: $\Delta \Q^{(2)}_\text{loc}$ (solid blue line), $\Delta \Q^{(2)}_\text{hop}$ (solid green line), $\Delta \Q^{(2)}_\text{nl}$ (long-dashed red line). The contribution of $\Delta \Q^{(2e)}_\text{hop}$ is also shown.}
  \label{fig:Fig6}
\end{figure}

This leads us to investigate the reason why $\Delta\Q^{(e2)}_\text{hop}>0$. To this end, we plot in Fig.~\ref{fig:Fig6} the quantities defined in Eqs.~\eqref{eq:DeltaFlux} and \eqref{eq:DeltaFluxContribution} involving atom $2$ at stationarity normalized by $P$, the flux of energy pumped into the system. In the interval of $d$ in which efficiency amplification is realized, the local flux of atom $2$ is clearly enhanced, meaning that the presence of pumping increases the energy drawn locally by atom $2$ from the bath, reaching a maximum of $\sim 15\,P$ close to $d\simeq1\,\mu$m. The non-local contribution $\Delta \Q^{(2)}_\mathrm{nl}$ is also increased, but is constant and relatively small compared to $\Delta \Q^{(2)}_\mathrm{loc}$ in the efficiency-amplification region.

On the contrary, the hopping fluxes show a behavior opposed to the one of $\Delta \Q^{(2)}_\mathrm{loc}$, highlighting the fact that the energy absorbed by atom $2$ from the environment is almost entirely transmitted by hopping to the rest of the chain. Moreover, as illustrated by $\Delta \Q^{(2e)}_\text{hop}$, most of this energy is yielded to the extraction site $e$.

The previous observations illustrate the efficiency-amplification mechanism: in the presence of pumping, as the extraction site is moved away from the rest of the chain, the energy absorbed by atom $2$ from the EM bath is significantly increased with respect to the no-pumping scenario. Most of this energy is transmitted to atom $e$ through hopping flux and results in an efficiency $\chi \geq 1$, meaning that the additional energy extracted from the chain is greater than the one injected in site $p$.

\begin{figure}[h!]
  \centering
  \includegraphics[width=.48\textwidth]{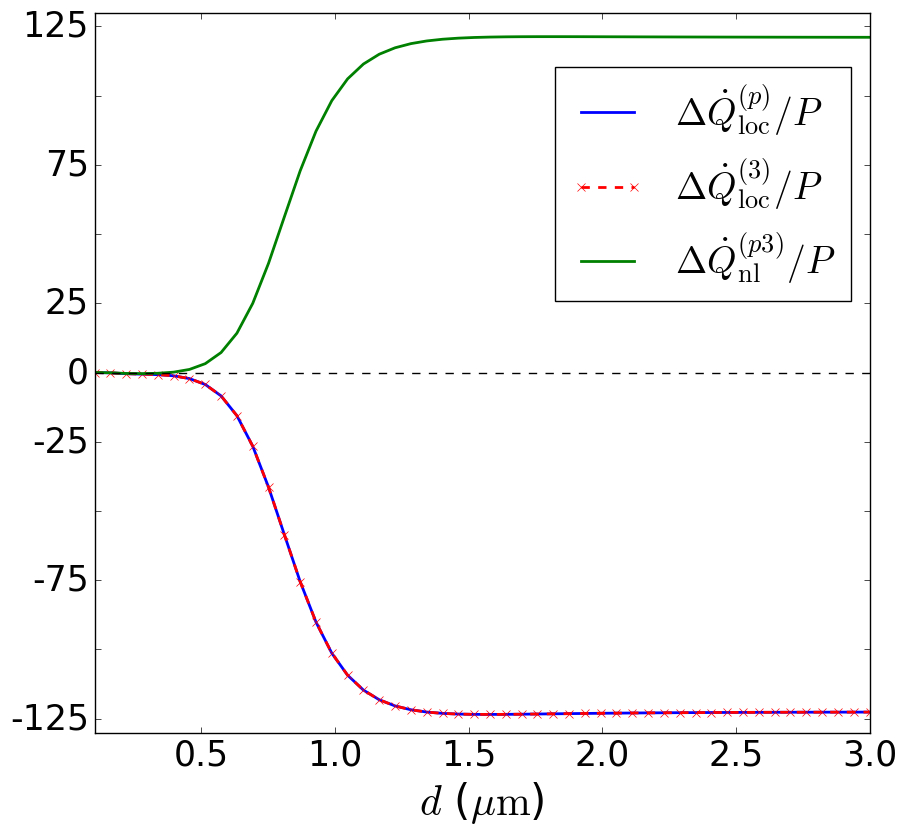}
  \caption{Difference between the pumping and no-pumping cases of the local heat fluxes of the atoms $p$ ($\Delta\Q_\text{loc}^{(p)}$, solid blue line) and 3 ($\Delta\Q_\text{loc}^{(3)}$, dashed red line with crosses), and of the non-local flux between them ($\Delta\Q_\text{nl}^{(p3)}$, solid green line) as a function of the displacement of $e$. These quantities are normalized by the energy-flux pumped into the system.}
  \label{fig:Dissipative_fluxes_p3}
\end{figure}

The behavior of the local flux of atom 2 being modified in the presence of pumping suggests that the others atoms might be subject to a change in their interaction with the EM bath as well. To explore this possibility, we plot in Fig.~\ref{fig:Dissipative_fluxes_p3} the difference of the local and non-local heat fluxes of atoms $p$ and 3 between the scenarios with and without pumping, as a function of the displacement of $e$. As $d$ increases, both atoms experience a similar diminution of their local fluxes which reaches a constant value when $e$ is far enough. On the other hand, $\Delta\Q_\text{nl}^{(p3)}$ has an analogous behavior but with an opposite sign. As a consequence, the energy loss occurring through the local channels $\Q_\text{loc}^{(p)}$ and $\Q_\text{loc}^{(3)}$ is almost entirely counterbalanced by the absorption of excitations through their non-local flux.

The analysis of the fluxes performed this far gives us an interpretation of the efficiency-amplification mechanism. Pumping excitations into the chain results in a reorganization of the interactions between the atoms and the environmental EM bath, to the benefit of the efficiency. More specifically, the triplet $(p,2,3)$ plays the role of an excitation injector: the collective interaction of the couple $(p,3)$ has the effect of singling out the atom 2, which sees its locally-drawn energy improved. The energy absorbed by the triplet is then transmitted to the extraction site $e$ via hopping, with notably a particularly enhanced flux of the pair $(2,e)$. As a result, the energy extracted from $e$ is not only greater than in the no-pumping scenario, but of an amount which can go largely beyond the energy injected in the chain.

According to this interpretation, the non-monotonic behavior of the efficiency as a function of $d$ becomes clear. On the one hand, the excitation injector needs to be relatively isolated from the rest of the system, otherwise the symmetry of this triplet is broken, resulting in a different heat fluxes distribution and thus in the disappearance of the effect. On the other hand, if the triplet $(p,2,3)$ is too far away from the rest of the chain, the hopping fluxes will be diminished, resulting in a lowered efficiency. Therefore, the optimal displacement of $e$ stems from the best compromise between the confinement of the excitation injector and its proximity to the extraction site.

Having so far performed the investigation of the efficiency amplification with respect to different parameters and of the mechanism producing this effect, both of them at steady state, we dedicate the following section to the dynamics of the energy transport occurring in our system. This will give us a deeper understanding of the process producing an amplified efficiency with, in particular, the identification of the establishment over time of the excitation injector.

\subsection{Dynamics}
\label{Dynamics}

In this subsection, we are interested in the dynamics of the system previously studied which, at stationarity, produces long-range efficiency amplification. Throughout this part, the distance of the extraction site from atom 3 is $d=1.03\,\mu$m, resulting in an efficiency of $\chi=10.21$.

\begin{figure}[h!]
  \centering
  \includegraphics[width=.48\textwidth]{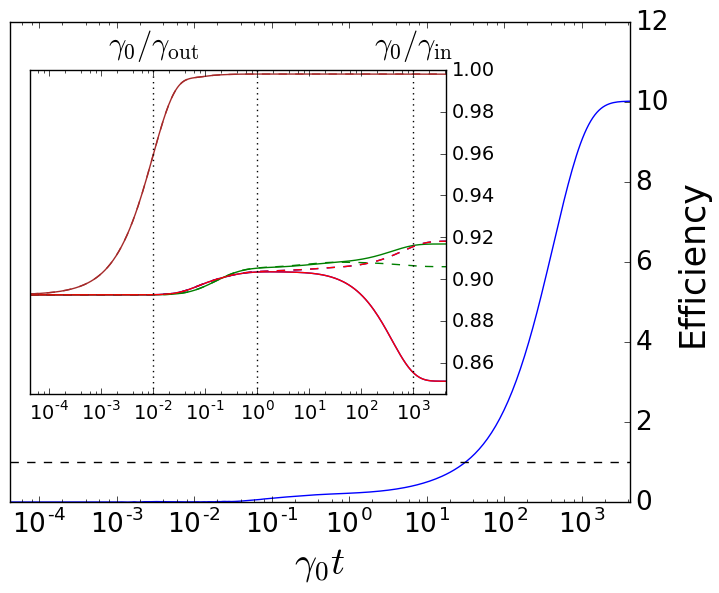}
  \caption{Time evolution of the efficiency $\chi$ (solid blue line). Inset: ground state population of each atom (index 0 corresponds to the no-pumping scenario): $p^{(p)}$ (solid red line), $p^{(p)}_0$ (dashed red line), $p^{(2)}$ (solid green line), $p^{(2)}_0$ (dashed green line), and $p^{(e)}$ (solid brown line). The curves $p^ {(3)}$, $p^ {(3)}_0$ and $p^{(e)}_0$ overlap with $p^ {(p)}$, $p^ {(p)}_0$ and $p^ {(e)}$, respectively.}
  \label{fig:Fig7}
\end{figure}

First, to have an overview of the dynamics of the system, let us begin this analysis with Fig.~\ref{fig:Fig7} showing the time evolution of the efficiency, as well as the ground state population of each atom of the chain. It comes clear from the plots that the latter evolves according to three different time scales. The first one is due to the extraction, characterized by the rate $\gamma_\text{out}$, which occurs very early and drives the population of the atom $e$ close to $p^{(e)}\simeq 1$ similarly in both cases, i.e. with and without pumping ($p^{(e)}=p^{(e)}_0$, the index 0 referring to quantities in the absence of pumping).

The second time scale entering into play triggers the evolution of the population of atoms $p$, $2$ and $3$ and is described by the local ($\gamma_j$) and non-local ($\gamma_{jk}$) rates. These are of the order of magnitude of $\gamma_0$, which is the only one represented in the inset of Fig.~\ref{fig:Fig7}. It is worth stressing that the ground state population of atoms $p$ and $3$ have the same evolution whether the pumping is performed ($p^{(p)}=p^{(3)}$) or not ($p^{(p)}_{0}=p^{(3)}_{0}$), and that in both scenarios, this evolution is different from the one of the ground state population of atom 2.

Besides, the populations of each atom have a different dynamics between the two scenarios $p^{(j)}\neq p^{(j)}_0 $ (except for the atom $e$, as mentioned before). Not surprisingly, this difference becomes significant only at the third time scale of the system, determined by the pumping rate $\gamma_\text{in}$. In particular, the stationary ground state populations of atoms $p$ and $3$ in the no-pumping scenario are greater than the one of atom $2$ ($p^{(p)}_0=p^{(3)}_0>p^{(2)}_0$) while the situation is reversed in the pumping case ($p^{(2)}>p^{(p)}=p^{(3)}$). In other words, the presence of pumping has a notable effect on the population distribution of these atoms, which suggests why the efficiency is increasing the most at this stage of the evolution.

\begin{figure}[h!]
  \centering
  \includegraphics[width=.48\textwidth]{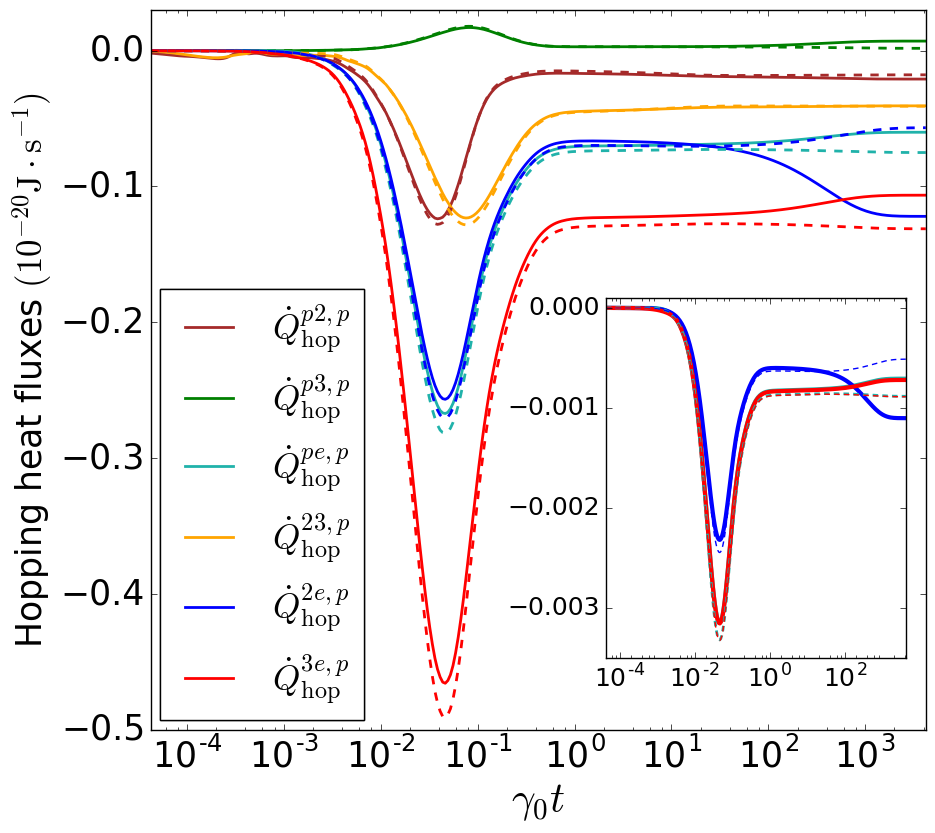}
  \caption{Time evolution of the hopping fluxes for each couple of atoms in the presence of pumping. Inset: Imaginary part of the coherences of couples ($p$,$e$), (2,$e$) and (3,$e$) also in the presence of pumping, with the same color scheme. The curves of pairs ($p$,$e$) and (2,$e$) are superimposed. The dotted lines, both in the main part of the figure and in the inset, represent the same quantities in the absence of pumping.}
  \label{fig:Fig8}
\end{figure}

In order to explain this modification in the distribution of the ground state populations, we now turn our investigation toward the dynamics of the heat fluxes. We start with Fig.~\ref{fig:Fig8}, depicting the time evolution of the hopping ones. According to Eq.~\eqref{eq:HoppingFluxes2}, the dynamics of these energy exchanges is driven by the one of the imaginary part of the coherences, as confirmed by the inset.

\begin{figure}[h!]
  \centering
  \includegraphics[width=.48\textwidth]{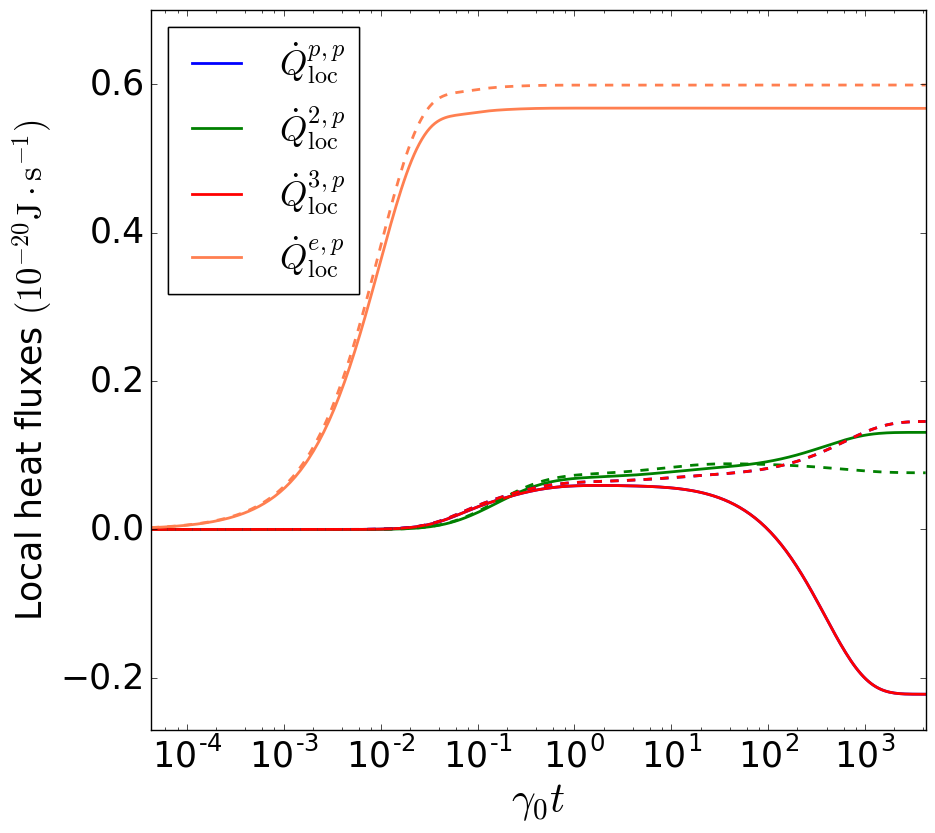}
  \caption{Time evolution of the local heat fluxes of each atom of the chain in the presence of pumping. The dotted lines are the same quantities in the absence of pumping (same color code).}
  \label{fig:Fig9}
\end{figure}

\begin{figure}[h!]
  \centering
  \includegraphics[width=.48\textwidth]{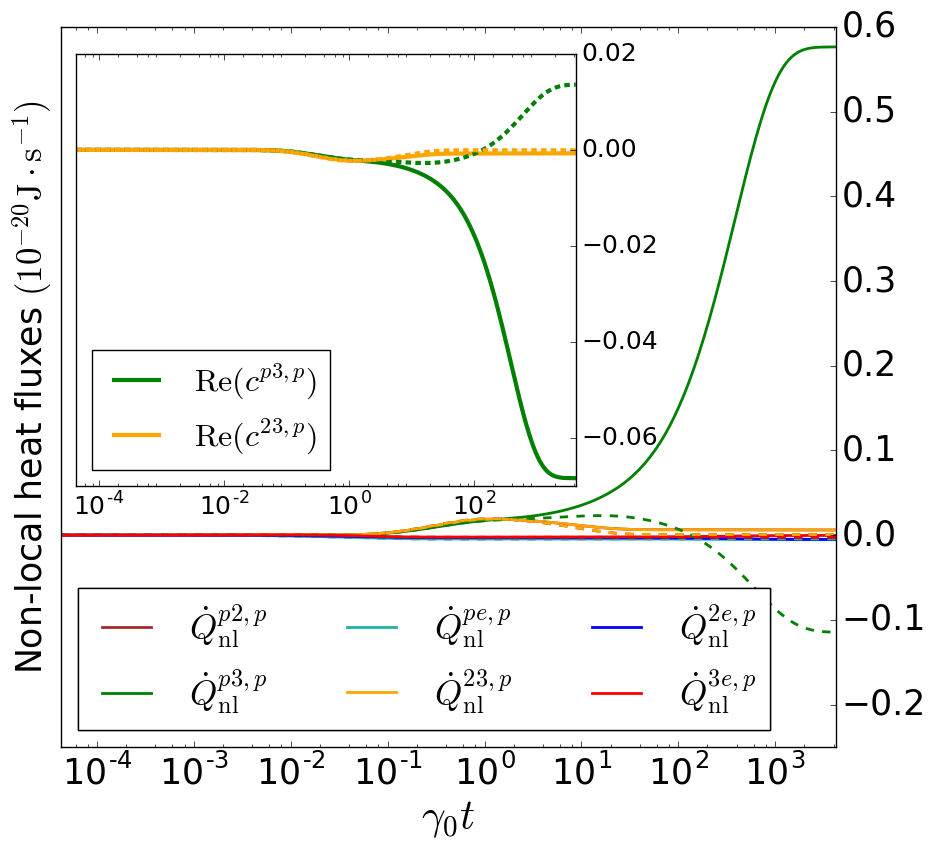}
  \caption{Dynamics of the non-local fluxes for each pair of atoms composing the chain in the pumping scenario. Inset: Real part of the coherences of the couples ($p$,3) and (2,3). The dotted lines are the same quantities in the absence of pumping (same color code).}
  \label{fig:Fig10}
\end{figure}

The peaks appearing in the interval $[10^{-2}, 10^{-1}]$ stem from the fact that the atom $e$ is in its ground state ($p^{(e)}\simeq 1$) due to extraction. As a consequence, coherences raise between subparts of the atomic system in order to draw it closer to its Gibbs state, thus producing hopping fluxes within the chain. Note that the ones with the greatest amplitudes are the fluxes involving atom $e$. The decrease of these peaks coincides with the appearance of the local and non-local heat fluxes whose dynamics is shown in Figs.~\ref{fig:Fig9} and \ref{fig:Fig10}.

Concerning the local heat fluxes, note that $\Q^{(e)}_\text{loc}$ begins its evolution earlier than the rest of the others because of the extraction. Moreover, $\Q^{(e)}_\text{loc}>0$ implies that this flux competes with the loss of energy induced by $\gamma_\text{out}$. Later in the evolution, the local and non-local fluxes of the other atoms enter into play. Note that only the couples $(p,3)$ and $(2,3)$ produce non-local contributions, and that they are driven by the real part of the coherences (inset of Fig.~\ref{fig:Fig10}).

The main distinction between the pumping and no-pumping scenarios occurs at the third time scale. The local fluxes of atoms $p$ and $3$ become negative, meaning that they are both loosing energy. However, the non-local flux stemming from the collective interaction between them increases significantly in the pumping case, such that the resulting flux between the field and this pair of atoms is positive: $\Q^{(p3)}_\text{nl}+\Q^{(p)}_\text{loc}+\Q^{(3)}_\text{loc}>0$. In other words, the subsystem $(p,3)$ is absorbing more energy than in the no-pumping scenario, which explains why their ground state population is lowered.

Besides, the local flux of atom $2$ increases with respect to the no-pumping case. Yet, its ground-state population is the closest one to $p^{(e)}$ and the imaginary parts of the coherences of the pair $(2,e)$ are also increased, inducing an enhanced hopping flux between these two atoms. On the contrary, the amplitudes of $\Q^{(pe)}_\text{hop}$ and $\Q^{(3e)}_\text{hop}$ are lowered, but remain positive, meaning that atoms $p$ and $3$ keep yielding energy to the extraction site (see Fig.~\ref{fig:Fig8}). This analysis suggests a complicated three-body coupled dynamics, in which two symmetric atoms build up correlations leaving to the central atom more freedom to strongly interact with the thermal environment. The description in terms of physical quantities pertaining to subparts of such a three-body system, if on the one hand provides a direct and clearer physical picture, on the other hand hinders the full complexity of this phenomenon. A complete study of the complicated structure of the interplay between atom-atom coherences and collective atoms-environment coupling goes however beyond the goal of this work.

\section{More atoms}
\label{sec:MoreAtoms}

Having studied in detail the 4-atom configuration, it has become clear that the excitation injector, i.e the triplet of the first three atoms of the chain, is a key ingredient to observe long-range efficiency amplification. According to Fig.~\ref{fig:Fig3}, the presence of additional atoms in the neighborhood of this triplet does not produce an enhanced efficiency. This happens because the geometry of the chain is different, inducing a redistribution of the heat fluxes which results in a break of the effect stemming from the excitation injector.

\begin{figure}[h!]
  \centering
  \includegraphics[width=.48\textwidth]{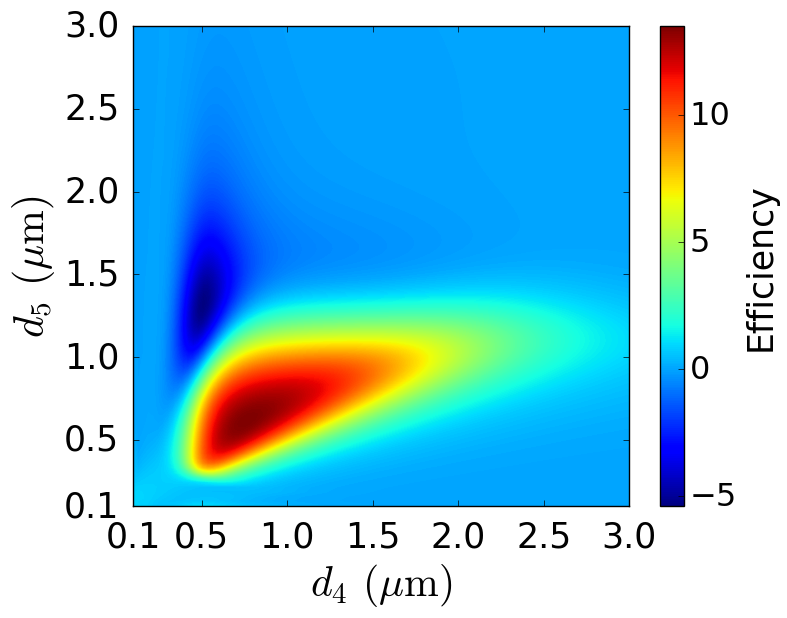}
  \caption{Density plot of the efficiency as a function of the displacement of atoms 4 and $e$ in a 5-atom linear chain. The other atoms are regularly separated by a distance $a=0.1\,\mu$m.}
  \label{fig:Fig11}
\end{figure}

With this information in mind, we have explored geometrical configurations with $N$ atoms in which the effect of the excitation injector is not killed. To this end, we allowed displacements not only of atom $e$, but also of any atom which does not belong to the triplet $(p,2,3)$. In the following simulations, the frequency is $\omega=1\times10^{14}\,\text{rad}\cdot\text{s}^{-1}$ and the temperature $T=361\,$K.  Figure~\ref{fig:Fig11} is a density plot of the efficiency as a function of the displacement of both atoms $4$ and $e$ in a 5-atom chain. There is clearly a region where long-range efficiency amplification is realized. Moreover, Tab.~\ref{tab:theTab} shows maxima of $\chi$ after displacement of the atoms that do not belong to the excitation injector in the case of $N$-atom chains ($N\in\{5,6,7\}$). They have been computed with the use of a genetic algorithm~\cite{haupt_practical_2004}. In the cases $N=4,5$, these maxima are global while they are local for the configurations with $N=6,7$ atoms. As one can see, for $N>4$ the long-range efficiency amplification is even higher than the 4-atom configuration. In particular, in the case of the 7-atom chain, the efficiency can reach values up to $1390\%$, which is $\sim1.4$ times greater than the maximum of the 4-atom chain ($1000\%$). In addition, the transport is performed over longer distances.

Finally, the presence of multiple local maxima of the efficiency (for $N=6,7$) indicates that these can be produced in different configurations of the chain when the number of atoms is large enough, thus providing geometrical flexibility to obtain an enhanced energy transport.

\begin{center}
  \begin{table}
  \begin{tabular}{|c|K{1.3cm}|K{1.3cm}|K{1.3cm}|K{1.3cm}|K{1.3cm}|}
    \hline
    $N$ & $d_4 (\mu \text{m})$ & $d_5 (\mu \text{m})$ & $d_6 (\mu \text{m})$ & $d_7 (\mu \text{m})$ & $\chi$ \\
    \hline
    \hline
    5 & 0.763 & 0.618 & \cellcolor{ashgrey} & \cellcolor{ashgrey} & 13.518 \\
    \hline
    \hline
    \multirow{2}{1cm}{\centering 6} & 0.556 & 1.076 & 0.578 & \cellcolor{ashgrey} & 13.982 \\
    \cline{2-3} \cline{4-4} \arrayrulecolor{ashgrey} \cline{5-5} \arrayrulecolor{black} \cline{6-6} & 0.782 & 0.564 & 0.518 & \cellcolor{ashgrey} & 13.631 \\
    \hline
    \hline
    7 & 0.768 & 0.581 & 0.380 & 0.419 & 13.908 \\
    \hline
  \end{tabular}
  \caption{Maxima of the efficiency for linear chains with different number of atoms. These maxima have been obtained with a genetic algorithm (GA) and are either global ($N=4,5$) or local ($N=6,7$). In each simulation, the variables are the displacement $d_j$ along the $x$-axis of the $j$-th atom with respect to its regular position, and such that  $j\notin(p,2,3)$. These three atoms are regularly distributed with a separation of $a=0.1\,\mu$m. Concerning the parameters used for the GA, the initial population  of each iteration was $\text{p}_\text{GA}=10^3$ with only the best half of it surviving. The mutation rate was fixed at $r_\text{GA}=20\%$, and convergence has been reached when the relative difference between each couple of the first 20 positions as well as the difference between their efficiencies were at most $10^{-3}$, except for $N=7$ for which it was $10^{-2}$. }
  \label{tab:theTab}
  \end{table}
\end{center}

\section{Conclusions}
\label{sec:Conclusions}

We studied energy transport within a chain of two-level quantum emitters weakly coupled to an environmental blackbody radiation. In particular, we investigated the efficiency of energy transport when excitations are incoherently pumped from one edge of the chain and extracted at the other one. As the main result, we have highlighted the remarkable appearance of a strongly amplified efficiency occurring for long-range transport.

More specifically, we have shown that this phenomenon is produced for non-regular configurations at finite temperature and results in an efficiency greater than the one observed with a regular-chain at low $T$. In particular, efficiency can reach values far greater than $100\%$ (e.g. $\chi\simeq1400\%$), meaning that the chain harvests additional energy from the bath during the transport process. Moreover, the length of the chain producing this amplified efficiency is remarkably greater than the regular configuration, thus allowing energy transport over longer distances.

We have also investigated, in the particular case of a 4-atom chain, the robustness of efficiency amplification with respect to relevant parameters of the system, such as the temperature, the atomic frequency and the distance of the extraction site. We have shown that this effect can be produced for a relatively wide range of temperatures and distances.

The investigation of the heat fluxes both at steady state and during the time evolution of the atomic system has provided an interpretation for the mechanism producing this amplification: the elementary block composed of the first three atoms of the chain plays the role of an excitation injector by absorbing energy from the environment. More specifically, the atoms ($p$,3) collectively draw excitations through non-local flux, while in addition atom 2 absorbs energy locally. The energy absorbed by the triplet is transmitted to the extraction site mainly through hopping fluxes.

We have also explored systems composed of $N=5,6,7$ atoms, where we provided specific examples of configurations producing a long-range efficiency amplification. Two encouraging results stem from this study when compared to the more elementary 4-atom chain: the efficiency can reach higher values ($\sim1.4$ times greater) and the chains producing such high-efficiency also transport energy over distances larger than the regular configuration.

Further investigations with an even larger number of atoms would be insightful, notably in the case of linear chains composed of multiple excitation injectors. Besides, the long-range efficiency amplification discussed in this paper could be possibly experimentally observed with, e.g., Rydberg atoms~\cite{saffman_2010} or quantum dots~\cite{komiyama_single-photon_2000,kammerer_mid-infrared_2005,homeyer_resonant_2009,barreiro_2012,wasserman_2009} playing the role of the quantum emitters.

\section*{ACKNOWLEDGEMENTS}
Authors acknowledge financial support from the Julian Schwinger Foundation.

\bibliographystyle{apsrev4-1}
\bibliography{References}

\end{document}